\documentclass{Interspeech2024}

\usepackage{cite}
\usepackage{booktabs}
\usepackage{multirow}
\usepackage{cleveref}

\interspeechcameraready

\title{Efficient Audio Captioning with Encoder-Level Knowledge Distillation}

\name[affiliation={1}]{Xuenan}{Xu}
\name[affiliation={2}]{Haohe}{Liu}
\name[affiliation={1}]{Mengyue}{Wu}
\name[affiliation={2}]{Wenwu}{Wang}
\name[affiliation={2}]{Mark D.}{Plumbley}

\address{
  $^1$MoE Key Lab of Artificial Intelligence X-LANCE Lab, Shanghai Jiao Tong University, China\\
  $^2$Centre for Vision, Speech and Signal Processing (CVSSP), University of Surrey, UK}
\email{\{wsntxxn, mengyuewu\}@sjtu.edu.cn, \{haohe.liu, w.wang, m.plumbley\}@surrey.ac.uk}


\begin{document}

\maketitle

\begin{abstract}
    
Significant improvement has been achieved in automated audio captioning (AAC) with recent models.
However, these models have become increasingly large as their performance is enhanced.
In this work, we propose a knowledge distillation (KD) framework for AAC.
Our analysis shows that in the encoder-decoder based AAC models, it is more effective to distill knowledge into the encoder as compared with the decoder.
To this end, we incorporate encoder-level KD loss into training, in addition to the standard supervised loss and sequence-level KD loss.
We investigate two encoder-level KD methods, based on mean squared error (MSE) loss and contrastive loss, respectively.
Experimental results demonstrate that contrastive KD is more robust than MSE KD, exhibiting superior performance in data-scarce situations.
By leveraging audio-only data into training in the KD framework, our student model achieves competitive performance, with an inference speed that is 19 times faster\footnote{An online demo is available at \url{https://huggingface.co/spaces/wsntxxn/efficient_audio_captioning}}.
\end{abstract}

\section{Introduction}
Automated audio captioning (AAC) is a cross-modal translation task that bridges the modalities of audio and text, aiming to generate textual descriptions for given audio inputs.
Recent advancements have shown significant improvements to the performance of the captioning models in accuracy~\cite{zhang2023actual,wu2023beats,xie2023enhance}, diversity~\cite{mei2022diverse} and generalizability~\cite{kim2023prefix}.
The popularity of the Detection and Classification of Acoustic Scenes and Events (DCASE) challenges has also attracted many researchers to contribute to the development of this field.

\begin{figure}
    \centering
    \includegraphics[width=1.0\linewidth]{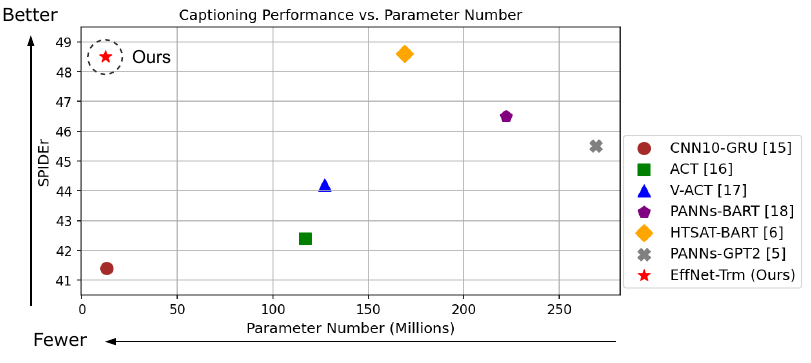}
    \caption{The comparison of performance-size tradeoff between our model and previous methods, evaluated on AudioCaps.}
    \label{fig:performance_vs_param}
\end{figure}

Improvements in AAC performance are often achieved at the cost of increased model complexity.
The state-of-the-art (SOTA) models~\cite{mei2023wavcaps,wu2023beats} typically employ deep convolutional neural networks (CNNs) or Transformers to extract embeddings from audio inputs, along with large-scale pre-trained Transformer models (e.g., BART~\cite{lewis2020bart}) for text generation.
Although these large-scale models achieve superior performance, they often require substantial computational overhead, memory and storage, posing challenges for their deployment on resource-constrained devices.
For example, HTSAT-BART~\cite{mei2023wavcaps} contains about 170 million parameters and requires 160 giga float-point-operations (FLOPs) for inference on a 10-second audio clip.
In addition, large-scale models are often over-parameterized for their target tasks (shown in \Cref{fig:performance_vs_param}).
However, to the best of our knowledge, model compression within the realm of AAC has attracted little attention.

Efforts have been made to reduce the computation cost by compressing models in speech and audio processing~\cite{liu2023multi,singh2023efficient,liu2023extremely}.
Knowledge distillation (KD), pruning and quantization have been used to develop small student models from large teacher models, by removing redundant connections and reducing parameter precision.
Despite advances in compressing classification models, the effectiveness of compression techniques on generation tasks, especially with the encoder-decoder framework, is rarely investigated.
To our knowledge, the only work is \cite{sridhar2023parameter} on developing parameter-efficient captioning models.
However, contrastive language audio pre-training (CLAP)~\cite{wu2023large} was utilized by \cite{sridhar2023parameter}, as a result, the involved parameters and computation cost remain substantial.

In this paper, we aim to fill the gap of model compression within AAC.
Specifically, we focus on KD since KD allows for the architecture flexibility that highly efficient student models can be used regardless of the teacher's architecture.
First, we analyze the bottleneck of distilling the encoder-decoder captioning framework widely adopted in AAC.
Our preliminary result shows that using a compact encoder results in a larger performance drop than using a compact decoder.
This reveals that the key to effective compression lies in developing an efficient and effective encoder.
Therefore, in our KD approach, we leverage audio embeddings from the teacher encoder for supervision (called encoder-level distillation).
Compared with previous works focusing on distilling classification models, the proposed encoder-level distillation provides an effective constraint on encoder outputs for distilling encoder-decoder AAC frameworks.

\begin{figure*}
    \centering
    \includegraphics[width=1.0\linewidth]{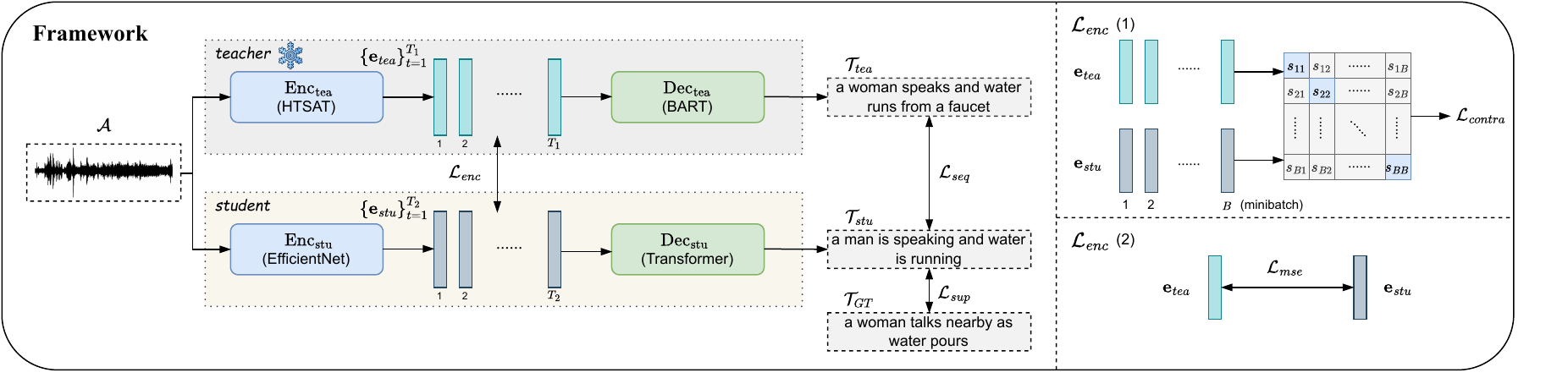}
    \caption{An overview of our proposed audio captioning knowledge distillation framework, which combines supervised loss $\mathcal{L}_{sup}$, sequence-level distillation loss $\mathcal{L}_{seq}$ and encoder-level distillation loss $\mathcal{L}_{enc}$ for training. We explore two kinds of $\mathcal{L}_{enc}$: 1) contrastive loss; 2) MSE loss.}
    \label{fig:approach}
\end{figure*}

We investigate two kinds of loss functions for this encoder-level distillation.
The first is the standard mean squared error (MSE) loss ($\text{KD}_\text{mse}$), aiming at minimizing the distance between the student and the teacher audio embeddings in $L^2$ space.
The second is a contrastive loss ($\text{KD}_\text{contra}$), where embeddings of the same audio clip obtained from teacher and student encoders are brought closer while embeddings from different audio clips are pushed further apart.
With this design, we encourage the student encoder to learn the same ability as the teacher to distinguish between different audio clips.
Based on the EfficientNet~\cite{tan2019efficientnet} architecture, we combine the standard sequence level KD and our proposed encoder-level KD for training.
Experimental results show that $\text{KD}_\text{contra}$ performs more robustly than $\text{KD}_\text{mse}$ in the data scarcity scenario.
Compared with training from scratch, KD achieves significant performance improvement.
Incorporating unannotated audio-only data into training further improves the performance, resulting in an efficient captioning model.
Our model achieves comparable performance to its teacher model, which is nearly SOTA, with only 6.5\% of the parameters.
Comparison of our model and previous models~\cite{xu2021investigating,mei2021audio,liu2022visually,gontier2021automated,mei2023wavcaps,kim2023prefix} in \Cref{fig:performance_vs_param} indicates the effectiveness and efficiency of our model.

\section{Proposed Knowledge Distillation Framework}

\subsection{Framework Overview}

\Cref{fig:approach} is an overview of our proposed audio captioning KD framework.
For an audio clip $\mathcal{A}$ and the corresponding caption $\mathcal{T}_{GT}$, the encoders $\mathrm{Enc}(\cdot)$ transform $\mathcal{A}$ into audio embedding sequences:
\begin{align}
    \begin{split}
    \{\mathbf{e}^t_{tea}\}_{t=1}^{T_1} &= \mathrm{Enc}_{tea}(\mathcal{A})\\
    \{\mathbf{e}^t_{stu}\}_{t=1}^{T_2} &= \mathrm{Enc}_{stu}(\mathcal{A})
    \end{split}
\end{align}
where $T_1$ and $T_2$ are sequence lengths or embeddings from the teacher and student encoders since they may use different temporal resolutions.
Then the teacher model predicts the caption $\mathcal{T}_{tea}$ conditioned on the encoded embeddings:
\begin{equation}
    \mathcal{T}_{tea} = \mathrm{Dec}_{tea}(\{\mathbf{e}^t_{tea}\}_{t=1}^{T_1}).
\end{equation}
where $\mathcal{T}_{tea}$ and $\mathcal{T}_{GT}$ are both utilized as supervision signals for the training of the student model.
Taking $\mathcal{T}_{GT}$ as conditions, the student model is trained by minimizing the Kullback-Leibler (KL) divergence between the predicted word distribution and the ground truth distribution as follows,
\begin{align}
    \begin{split}
        p^{GT} &= \mathrm{Dec}_{stu}(\{\mathbf{e}^t_{stu}\}_{t=1}^{T_2}, \mathcal{T}_{GT})\\
        \mathcal{L}_{sup} &= -\sum_{n=1}^{N_1}\log(p^{GT}_{n, (\mathcal{T}_{GT})_n})
    \end{split}
\end{align}
where $p^{GT} \in \mathbb{R}^{N_1 \times \left| \mathcal{V} \right|}$ is the predicted word probability.
$N_1$ is the word number of the ground truth caption and $\mathcal{V}$ is the vocabulary.
$(\mathcal{T}_{GT})_n$ denotes the index of the $n$-th word of the caption so $p^{GT}_{n, (\mathcal{T}_{GT})_n}$ is the predicted probability of the $n$-th ground truth word.
Similarly, $\mathcal{T}_{tea}$ is utilized to calculate the sequence-level KD loss:
\begin{align}
    \begin{split}
        p^{tea} &= \mathrm{Dec}_{stu}(\{\mathbf{e}^t_{stu}\}_{t=1}^{T_2}, \mathcal{T}_{tea})\\
        \mathcal{L}_{seq} &= -\sum_{n=1}^{N_2}\log(p^{tea}_{n, (\mathcal{T}_{tea})_n}).
    \end{split}
\end{align}
We use a tokenizer with a smaller vocabulary size for the student model since the tokenizer of the teacher model involves a large number of parameters.
As a consequence, word probabilities predicted by the teacher model given the ground truth caption cannot be used for training.
In addition to these standard KD losses, we add a constraint on the audio encoder output ($\mathcal{L}_{enc}$), which will be elaborated in \Cref{subsec:encoder_kd}.
The final training loss is the combination of losses from different levels:
\begin{equation}
    \label{eq:total_loss}
    \mathcal{L} = \mathcal{L}_{sup} + \mathcal{L}_{seq} + \mathcal{L}_{enc}.
\end{equation}


\paragraph*{Teacher}
We take the official HTSAT-BART checkpoint from \cite{mei2023wavcaps} as the teacher.
The encoder is a Swin-Transformer~\cite{liu2021swin} while the decoder takes the $\text{BART}_\text{base}$ architecture, consisting of 12 Transformer layers.
Although the teacher achieves competitive performance, the model is heavily parameterized, especially the deep decoder pre-trained on general natural language tasks.
\paragraph*{Student}
Motivated by the finding that Transformer serves as an effective teacher for CNN students~\cite{schmid2023efficient}, we adopt EfficientNet-B2~\cite{tan2019efficientnet} as the student encoder.
Depthwise convolution~\cite{chollet2017xception} is used instead of the standard convolution to enhance parameter efficiency.
For the decoder, we adopt a shallow 2-layer Transformer due to its competitive performance in DCASE challenges~\cite{xu2022sjtu}.
Such a combination reduces the parameter number from 170 million to 11 million.
We refer to our student model as ``EffNet-Trm''.

\subsection{Encoder-Level Knowledge Distillation}
\label{subsec:encoder_kd}
The comparison of learning difficulty for the encoder and decoder, which will be shown in \Cref{subsec:bottleneck_analysis}, reveals that the bottleneck of training a small student captioning model lies in training an efficient encoder.
Therefore, we add an extra constraint $\mathcal{L}_{enc}$ to guide the student encoder to generate embeddings that closely follow teacher-generated embeddings.
As the right part of \Cref{fig:approach} shows, two kinds of loss functions, $\mathcal{L}_{contra}$ and $\mathcal{L}_{mse}$, are investigated.
For both loss types, projection layers are utilized but we omit them for simplicity.

\subsubsection{Distillation via Contrastive Learning}
We first explore the contrastive loss which is widely utilized in self-supervised learning~\cite{oord2018representation}.
Mean pooling is applied to audio embedding sequences to obtain clip-level embeddings.
Then, these embeddings are projected to the same dimension by two projection layers:
\begin{equation*}
    \mathbf{e}_{tea} = \mathrm{Proj}_{tea}(\frac{1}{T_1}\sum_{t=1}^{T_1}\mathbf{e}_{tea}^t), 
    \mathbf{e}_{stu} = \mathrm{Proj}_{stu}(\frac{1}{T_2}\sum_{t=1}^{T_2}\mathbf{e}_{stu}^t).
\end{equation*}
We calculate the cosine similarity $s(i, j)$ between the teacher embedding of the $i$-th sample and the student embedding of the $j$-th sample in a minibatch.
The contrastive loss is defined as: 
\begin{align}
    \begin{split}
        &s(i, j) = \frac{\mathbf{e_{tea}}(i)\cdot\mathbf{e_{stu}}(j)^{\mathrm{T}}}{
        \Vert \mathbf{e_{tea}}(i) \Vert \cdot \Vert \mathbf{e_{stu}}(j) \Vert }\\
        &\mathcal{L}_{i,1} = -\log \frac{\exp\left(s(i, i\right) / \tau )}{\sum_{j=1}^B \exp(s\left(i, j\right) / \tau)}\\
        &\mathcal{L}_{i,2} = -\log \frac{\exp\left(s(i, i\right) / \tau )}{\sum_{j=1}^B \exp(s\left(j, i\right) / \tau)}\\
        &\mathcal{L}_{contra} = \frac{1}{B}\sum_{i=1}^B(\mathcal{L}_{i,1} + \mathcal{L}_{i,2})
    \end{split}
\end{align}
where $B$ is the batch size and $\tau$ is the scaling temperature.
$\mathcal{L}_{contra}$ is proposed to guide the student encoder to not only replicate the teacher encoder's outputs but also learn the underlying patterns that distinguish one audio sample from another.
Therefore, the student model is trained to generate distinct and effective embeddings for diverse audio inputs, which are desired by the decoder for accurate caption generation.

\subsubsection{Distillation via Optimizing Mean Squared Error}
$\mathcal{L}_{mse}$ is the standard embedding level loss used in KD.
After mean pooling and a projection layer, the $L_2$ distance between the teacher and student embedding is minimized:
\begin{align}
    \begin{split}
        \mathbf{e}_{tea} = \frac{1}{T_1}\sum_{t=1}^{T_1}&\mathbf{e}_{tea}^t, \mathbf{e}_{stu} = \mathrm{Proj}_{stu}(\frac{1}{T_2}\sum_{t=1}^{T_2}\mathbf{e}_{stu}^t)\\
        &\mathcal{L}_{mse} = \lVert \mathbf{e}_{tea} - \mathbf{e}_{stu} \rVert^2.
    \end{split}
\end{align}

Here in $\mathcal{L}_{mse}$, the student is trained to exactly follow the original teacher embedding so no projection is applied to teacher embeddings. 
The decoder uses $\{\mathrm{Proj}_{stu}(\mathbf{e}_{stu}^t)\}_{t=1}^{T_2}$ for inference so $\mathrm{Proj}_{stu}$ is used during inference.
In contrast, for $\mathcal{L}_{contra}$, the projection $\mathrm{Proj}_{stu}$ is only used during training, while for inference, the decoder still uses $\{\mathbf{e}_{stu}^t\}_{t=1}^{T_2}$.

\subsection{Training with Audio-Only Data}
With a strong teacher, we further leverage unannotated audio data to augment the training data.
The teacher is used to generate pseudo caption labels for audio data.
Therefore, the available training data is not limited to small-scale annotated audio-text pairs.
In practice, we use audio-only data that share the same distribution as the original dataset for augmentation to prevent domain mismatch induced by additional data.
For audio-only data, the loss function in \Cref{eq:total_loss} becomes $\mathcal{L}_{seq} + \mathcal{L}_{enc}$ since $T_{GT}$ is not available.

\section{Experimental Setup}

\subsection{Dataset}

In this work, we conduct experiments on Clotho~\cite{drossos2020clotho} and AudioCaps~\cite{kim2019audiocaps}.
AudioCaps is the largest human-annotated AAC dataset, containing over 50k audio-text pairs.
Since AudioCaps is a subset of AudioSet~\cite{gemmeke2017audio}, we use the whole AudioSet as the audio-only data.
Compared with AudioCaps, Clotho is a small-scale dataset with 6k audio clips.
Since Clotho originates from Freesound~\cite{font2013freesound}, we use Freesound as the audio-only data.
To reduce memory consumption during training, we only use audio clips shorter than 300 seconds.
A segment of 10 seconds is randomly cropped as the training sample.

\subsection{Hyper-parameters}

We use EfficientNet-B2 pre-trained on AudioSet to initialize the audio encoder of the student model.
The Transformer decoder is trained from scratch.
The whole model is trained for 25 epochs with a batch size of 32.
When audio-only data is incorporated into training, we use 16 original samples and 16 augmented ones in each iteration. 
We warm up the learning rate linearly to $5\times10^{-4}$ in the first 5 epochs and then exponentially reduce it to $5\times10^{-7}$.
Label smoothing with $\alpha = 0.1$ is used in $\mathcal{L}_{sup}$ and $\mathcal{L}_{seq}$ to smooth the ground truth distribution.
During inference, we adopt beam search with a beam size of 3.

\subsection{Evaluation Metrics}
For performance evaluation, traditional metrics, including BLEU, ROUGE, METEOR, CIDEr and SPICE~\cite{xu2023beyond} are used.
We also report a more advanced model-based FENSE, which shows a better correlation with human judgments.
To evaluate the size and memory footprint of our model, we also compare parameter numbers, FLOPs, and inference time of our model with the teacher.
\section{Results}

\subsection{Bottleneck Analysis}
\label{subsec:bottleneck_analysis}

We first investigate the bottleneck of KD in the encoder-decoder framework, i.e., which part, the encoder or the decoder, is more difficult to distill from the teacher model.
We initialize one part of the student model (encoder or decoder) with pre-trained parameters and freeze it while making the other part trainable, results compared in \Cref{tab:bottleneck_analysis}.
Row 1 shows the teacher's performance while in row 2, we set the encoder of the student model to be the frozen encoder of the teacher, and train the decoder from scratch.
\begin{table}[htpb]
    \centering
    \caption{Analysis on distillation bottleneck. ``EffNet'' and ``Trm'' denote the EfficientNet encoder and Transformer decoder in the student model. ``F'' means frozen.}
    \begin{tabular}{c|c|c|c}
    \toprule
    Encoder & Decoder & SPIDEr & FENSE \\
    \midrule
    HTSAT & BART & 48.6 & 64.2 \\
    HTSAT (F) & Trm & 48.8 & 63.6 \\
    EffNet & Trm (F) & 46.9 & 62.2 \\
    \bottomrule
    \end{tabular}
    \label{tab:bottleneck_analysis}
\end{table}

\begin{table*}[htpb]
    \centering
    \footnotesize
    \caption{Captioning performance of the distilled student, the teacher, and previous approaches. For student training, we report the mean and standard deviation of three random runs. We report the result of our implementation so there is a difference with the original literature for some approaches. ``Size'' denotes the model size measured in parameter numbers.}
    \begin{tabular}{c|l|l|r||lllllll}
    \toprule
    Dataset & \multicolumn{2}{c|}{Model} & Size / M & $\text{BLEU}_\text{4}$ & $\text{ROUGE}_\text{L}$ & METEOR & CIDEr & SPICE & FENSE \\
    \midrule
    \multirow{8}{*}{Clotho} & \multicolumn{2}{c|}{DCASE2023 baseline} & 98 & 17.1 & 38.5 & 17.8 & 41.3 & 11.9 & 46.8 \\
    & \multicolumn{2}{c|}{DCASE2023 winner~\cite{wu2023beats}} & 1368 (127) & - & - & 19.3 & 50.6 & 14.6 & 52.6 \\
    \cline{2-10}
    & \multicolumn{2}{c|}{Teacher~\cite{mei2023wavcaps}} & 169 & 17.3 & 38.7 & 18.7 & 47.8 & 13.4 & 51.8\\
    \cline{2-10}
    & \multirow{5}{*}{Student} & EffNet-Trm (scratch) & 11 & 17.1$_{\pm0.1}$ & \textbf{39.4$_{\pm0.2}$} & \textbf{18.7$_{\pm0.0}$} & 44.0$_{\pm0.2}$ & \textbf{13.1$_{\pm0.1}$} & 46.4$_{\pm0.3}$\\
    & & Proposed $\text{KD}_\text{mse}$ & 12 & 16.9$_{\pm0.2}$ & 38.6$_{\pm0.2}$ & 18.4$_{\pm0.1}$ & 44.0$_{\pm0.7}$ & 13.0$_{\pm0.1}$ & 48.7$_{\pm0.1}$\\
    & & \hspace{1em} + Audio-only Data & 12 & 17.5$_{\pm0.2}$ & 38.6$_{\pm0.0}$ & 18.4$_{\pm0.0}$ & 45.9$_{\pm0.3}$ & \textbf{13.1$_{\pm0.1}$} & 50.2$_{\pm0.4}$\\
    & & Proposed $\text{KD}_\text{contra}$ & 11 & \textbf{17.7$_{\pm0.1}$} & 38.9$_{\pm0.2}$ & 18.6$_{\pm0.0}$ & 45.4$_{\pm0.4}$ & 12.9$_{\pm0.2}$ & 49.3$_{\pm0.3}$\\
    & & \hspace{1em} + Audio-only Data & 11 & 17.7$_{\pm0.2}$ & 38.9$_{\pm0.0}$ & 18.5$_{\pm0.0}$ & \textbf{46.8$_{\pm0.1}$} & 13.0$_{\pm0.1}$ & \textbf{50.2$_{\pm0.0}$}\\
    \midrule
    \multirow{9}{*}{AudioCaps} & \multicolumn{2}{c|}{CNN10-GRU~\cite{xu2021investigating}} & 13 & 23.1 & 46.7 & 22.9 & 66.0 & 16.8 & 57.9 \\
    & \multicolumn{2}{c|}{ACT~\cite{mei2021audio}} & 117 & 25.2 & 48.1 & 23.3 & 67.9 & 16.8 & 60.2 \\
    & \multicolumn{2}{c|}{PANNs-BART~\cite{gontier2021automated}} & 222 & 26.6$_{\pm0.9}$ & 49.3$_{\pm0.4}$ & 24.1$_{\pm0.3}$ & 75.3$_{\pm0.9}$ & 17.6$_{\pm0.3}$ & - \\
    \cline{2-10}
    & \multicolumn{2}{c|}{Teacher~\cite{mei2023wavcaps}} & 169 & 28.5 & 50.7 & 25.0 & 79.0 & 18.2 & 64.2\\
    \cline{2-10}
    & \multirow{5}{*}{Student} & EffNet-Trm (scratch) & 11 & 27.7$_{\pm0.7}$ & 50.2$_{\pm0.2}$ & 24.5$_{\pm0.2}$ & 73.9$_{\pm1.2}$ & 18.1$_{\pm0.2}$ & 61.5$_{\pm0.2}$\\
    & & Proposed $\text{KD}_\text{mse}$ & 12 & 28.2$_{\pm0.3}$ & 50.8$_{\pm0.1}$ & 24.9$_{\pm0.1}$ & 78.6$_{\pm0.6}$ & 18.1$_{\pm0.2}$ & 63.3$_{\pm0.1}$ \\
    & & \hspace{1em} + Audio-only Data & 12 & \textbf{28.6$_{\pm0.3}$} & \textbf{51.0$_{\pm0.3}$} & \textbf{25.0$_{\pm0.1}$} & \textbf{78.8$_{\pm0.3}$} & 18.2$_{\pm0.1}$ & 63.6$_{\pm0.1}$ \\
    & & Proposed $\text{KD}_\text{contra}$ & 11 & 28.4$_{\pm0.5}$ & 50.7$_{\pm0.3}$ & 24.8$_{\pm0.1}$ & 77.8$_{\pm0.4}$ & 18.2$_{\pm0.1}$ & 63.5$_{\pm0.1}$ \\
    & & \hspace{1em} + Audio-only Data & 11 & 28.6$_{\pm0.5}$ & 50.8$_{\pm0.2}$ & 24.9$_{\pm0.1}$ & 78.5$_{\pm0.9}$ & 18.0$_{\pm0.2}$ & \textbf{63.7$_{\pm0.1}$}\\
    \bottomrule
    \end{tabular}
    \label{tab:result}
\end{table*}

Despite a small gap in FENSE, the student achieves competitive performance, as shown by a slightly higher SPIDEr.
In Row 3, we freeze the decoder as the pre-trained one from Row 2 and replace the HTSAT encoder with EfficientNet.
Compared with replacing the decoder, there is a larger performance drop in this case, indicating that a smaller encoder has a more significant impact than a smaller decoder.
Therefore, we place more emphasis on reducing the performance gap between the teacher encoder and the student encoder.

\subsection{Knowledge Distillation from the Teacher Model}
\Cref{tab:result} presents the effect of KD\footnote{For all KD settings, the improvement in FENSE is significant compared with training from scratch, with the corresponding $p$-value less than 0.05.}.
Our student model performs well even when trained from scratch.
On Clotho, in terms of some metrics (e.g., METEOR), the difference between the student and the teacher is small.
However, the most reliable metric FENSE shows a gap between teacher and student.
Compared with $\text{KD}_\text{mse}$, $\text{KD}_\text{contra}$ gives a larger improvement in CIDEr and FENSE.
The superior performance of $\text{KD}_\text{contra}$ on Clotho can be attributed to a smaller dataset size.
With limited training data, it is challenging for the student encoder to learn to replicate the teacher's embedding output.
However, the supervision of contrast between positive and negative pairs helps the student to discriminate between different audio inputs, aiding in learning the inherent patterns in various sound events and acoustic environments.
With audio-only data incorporated into training, the data scarcity problem is alleviated so that $\text{KD}_\text{mse}$ achieves similar performance to $\text{KD}_\text{contra}$.
On AudioCaps, the situation is similar since AudioCaps is large-scale.
With the combination of KD and audio-only data training, the student achieves significant improvement over training from scratch, especially on AudioCaps, where the student achieves comparable performance with the teacher.

Besides the teacher model, we also compare our student model with current well-performing models, which are mostly large in size.
For example, the top performing model in the DCASE2023 challenge incorporates 1368 million parameters, since Instructor-XL~\cite{su2022one}, which is a large model, is utilized for post-processing.
There are still 127 million parameters even without Instructor-XL.
In contrast, our EffNet-Trm achieves competitive performance with only about 10 million parameters, which is about 6\% of the teacher model.
Compared with the model with similar parameter numbers (e.g., CNN10-GRU), our model achieves much better performance.



\subsection{Inference Speedup}

We further compare the computation cost of the teacher and student on a resource-constrained device.
The FLOPs calculated in giga and inference latency on a Raspberry 4 Pi are shown in \Cref{tab:inference_speedup}.
We set the input as a 10-second audio clip and the predicted sequence length as 20 for both models.
With a compact and efficient architecture, the student model achieves speedup of a factor of about 20 compared to the teacher.
The FLOPs of the student model are only 2.3\% of the teacher's.

\begin{table}[htpb]
    \footnotesize
    \centering
    \caption{The latency on a Raspberry 4 Pi and giga FLOPs of the teacher and student. The inference is run 10 times and we report the average latency.}
    \begin{tabular}{c|c|c}
    \toprule
    Model & Latency / s & GFLOPs \\
    \midrule
    HTSAT-BART (teacher) & 45.2 & 160.7 \\
    EffNet-Trm (student) & 2.4 & 3.8 \\
    \bottomrule
    \end{tabular}
    \label{tab:inference_speedup}
\end{table}

\section{Conclusion}

In this paper, we have presented a teacher-student KD method for AAC to learn an efficient student model from a large-scale teacher model.
Our analysis reveals that for the encoder-decoder AAC framework, the key to KD is to learn an efficient encoder to extract representative audio embeddings.
Therefore, we combine the standard supervised loss and sequence-level KD loss with our proposed encoder-level KD loss for training.
We compare two types of encoder-level KD techniques, $\text{KD}_\text{mse}$ and $\text{KD}_\text{contra}$.
We further incorporate audio-only data to expand the training data.
Experimental results show that $\text{KD}_\text{contra}$ is more robust than $\text{KD}_\text{mse}$ in the data scarcity scenario but both methods achieve similar performance when sufficient training data is available.
Although with the limitation that there is still a gap between our model and current SOTA models in the data scarcity scenario, our student model achieves performance comparable to the teacher's with a speedup of $19\times$.

\section{Acknowledgements}
This work was supported in part by the British Broadcasting Corporation Research and Development (BBC R\&D), in part by Engineering and Physical Sciences Research Council (EPSRC) under Grant EP/T019751/1 "AI for Sound", and in part by a Ph.D Scholarship from the Centre for Vision, Speech and Signal Processing (CVSSP), Faculty of Engineering and Physical Science (FEPS), University of Surrey.
This work was also supported by Key Research and Development Program of Jiangsu Province~(No.BE2022059) and Guangxi major science and technology project~(No. AA23062062).
For the purpose of open access, the authors have applied a Creative Commons Attribution (CC BY) license to any Author Accepted Manuscript version arising.
This publication is supported by multiple datasets that are openly available at \cite{drossos2020clotho,kim2019audiocaps,mei2023wavcaps}.

\bibliographystyle{IEEEtran}
\bibliography{refs}

\begin{thebibliography}{10}
\providecommand{\url}[1]{#1}
\csname url@samestyle\endcsname
\providecommand{\newblock}{\relax}
\providecommand{\bibinfo}[2]{#2}
\providecommand{\BIBentrySTDinterwordspacing}{\spaceskip=0pt\relax}
\providecommand{\BIBentryALTinterwordstretchfactor}{4}
\providecommand{\BIBentryALTinterwordspacing}{\spaceskip=\fontdimen2\font plus
\BIBentryALTinterwordstretchfactor\fontdimen3\font minus \fontdimen4\font\relax}
\providecommand{\BIBforeignlanguage}[2]{{%
\expandafter\ifx\csname l@#1\endcsname\relax
\typeout{** WARNING: IEEEtran.bst: No hyphenation pattern has been}%
\typeout{** loaded for the language `#1'. Using the pattern for}%
\typeout{** the default language instead.}%
\else
\language=\csname l@#1\endcsname
\fi
#2}}
\providecommand{\BIBdecl}{\relax}
\BIBdecl

\bibitem{zhang2023actual}
Y.~Zhang, H.~Yu, R.~Du, Z.-H. Tan, W.~Wang, Z.~Ma, and Y.~Dong, ``{ACTUAL}: Audio captioning with caption feature space regularization,'' \emph{IEEE/ACM Trans. Audio, Speech Lang. Process.}, vol.~31, pp. 2643--2657, 2023.

\bibitem{wu2023beats}
S.-L. Wu, X.~Chang, G.~Wichern, J.-w. Jung, F.~Germain, J.~L. Roux, and S.~Watanabe, ``Beats-based audio captioning model with instructor embedding supervision and chatgpt mix-up,'' DCASE2023 Challenge, Tech. Rep., 2023.

\bibitem{xie2023enhance}
Z.~Xie, X.~Xu, M.~Wu, and K.~Yu, ``Enhance temporal relations in audio captioning with sound event detection,'' in \emph{Proc. ISCA Annu. Conf. Int. Speech Commun. Assoc.}, 2023, pp. 4179--4183.

\bibitem{mei2022diverse}
X.~Mei, X.~Liu, J.~Sun, M.~D. Plumbley, and W.~Wang, ``Diverse audio captioning via adversarial training,'' in \emph{Proc. IEEE Int. Conf. Acoust., Speech, Signal Process.}, 2022, pp. 8882--8886.

\bibitem{kim2023prefix}
M.~Kim, K.~Sung-Bin, and T.-H. Oh, ``Prefix tuning for automated audio captioning,'' in \emph{Proc. IEEE Int. Conf. Acoust., Speech, Signal Process.}, 2023, pp. 1--5.

\bibitem{mei2023wavcaps}
X.~Mei, C.~Meng, H.~Liu, Q.~Kong, T.~Ko, C.~Zhao, M.~D. Plumbley, Y.~Zou, and W.~Wang, ``Wavcaps: A chatgpt-assisted weakly-labelled audio captioning dataset for audio-language multimodal research,'' \emph{arXiv preprint arXiv:2303.17395}, 2023.

\bibitem{lewis2020bart}
M.~Lewis, Y.~Liu, N.~Goyal, M.~Ghazvininejad, A.~Mohamed, O.~Levy, V.~Stoyanov, and L.~Zettlemoyer, ``Bart: Denoising sequence-to-sequence pre-training for natural language generation, translation, and comprehension,'' in \emph{Proc. Annu. Meeting Assoc. Comput. Linguistics}, 2020, pp. 7871--7880.

\bibitem{liu2023multi}
Y.~Liu, H.~Sun, G.~Chen, Q.~Wang, Z.~Zhao, X.~Lu, and L.~Wang, ``Multi-level knowledge distillation for speech emotion recognition in noisy conditions,'' in \emph{Proc. ISCA Annu. Conf. Int. Speech Commun. Assoc.}, 2023, pp. 1893--1897.

\bibitem{singh2023efficient}
A.~Singh and M.~D. Plumbley, ``Efficient similarity-based passive filter pruning for compressing cnns,'' in \emph{Proc. IEEE Int. Conf. Acoust., Speech, Signal Process.}\hskip 1em plus 0.5em minus 0.4em\relax IEEE, 2023, pp. 1--5.

\bibitem{liu2023extremely}
B.~Liu, H.~Wang, and Y.~Qian, ``Extremely low bit quantization for mobile speaker verification systems under 1mb memory,'' in \emph{Proc. ISCA Annu. Conf. Int. Speech Commun. Assoc.}, 2023, pp. 1973--1977.

\bibitem{sridhar2023parameter}
A.~K. Sridhar, Y.~Guo, E.~Visser, and R.~Mahfuz, ``Parameter efficient audio captioning with faithful guidance using audio-text shared latent representation,'' \emph{arXiv preprint arXiv:2309.03340}, 2023.

\bibitem{wu2023large}
Y.~Wu, K.~Chen, T.~Zhang, Y.~Hui, T.~Berg-Kirkpatrick, and S.~Dubnov, ``Large-scale contrastive language-audio pretraining with feature fusion and keyword-to-caption augmentation,'' in \emph{Proc. IEEE Int. Conf. Acoust., Speech, Signal Process.}, 2023, pp. 1--5.

\bibitem{tan2019efficientnet}
M.~Tan and Q.~Le, ``Efficientnet: Rethinking model scaling for convolutional neural networks,'' in \emph{Proc. ICML}.\hskip 1em plus 0.5em minus 0.4em\relax PMLR, 2019, pp. 6105--6114.

\bibitem{xu2021investigating}
X.~Xu, H.~Dinkel, M.~Wu, Z.~Xie, and K.~Yu, ``Investigating local and global information for automated audio captioning with transfer learning,'' in \emph{Proc. IEEE Int. Conf. Acoust., Speech, Signal Process.}, 2021, pp. 905--909.

\bibitem{mei2021audio}
X.~Mei, X.~Liu, Q.~Huang, M.~D. Plumbley, and W.~Wang, ``Audio captioning transformer,'' in \emph{Proc. Detection Classification Acoust. Scenes Events}, 2021, pp. 211--215.

\bibitem{liu2022visually}
X.~Liu, Q.~Huang, X.~Mei, H.~Liu, Q.~Kong, J.~Sun, S.~Li, T.~Ko, Y.~Zhang, L.~H. Tang, M.~D. Plumbley, V.~Kılıç, and W.~Wang, ``Visually-aware audio captioning with adaptive audio-visual attention,'' in \emph{Proc. ISCA Annu. Conf. Int. Speech Commun. Assoc.}, 2023, pp. 2838--2842.

\bibitem{gontier2021automated}
F.~Gontier, R.~Serizel, and C.~Cerisara, ``Automated audio captioning by fine-tuning bart with audioset tags,'' in \emph{Proc. Detection Classification Acoust. Scenes Events}, 2021, pp. 170--174.

\bibitem{liu2021swin}
Z.~Liu, Y.~Lin, Y.~Cao, H.~Hu, Y.~Wei, Z.~Zhang, S.~Lin, and B.~Guo, ``Swin transformer: Hierarchical vision transformer using shifted windows,'' in \emph{Proc. IEEE/CVF Conf. Comput. Vis. Pattern Recognit.}, 2021, pp. 10\,012--10\,022.

\bibitem{schmid2023efficient}
F.~Schmid, K.~Koutini, and G.~Widmer, ``Efficient large-scale audio tagging via transformer-to-cnn knowledge distillation,'' in \emph{Proc. IEEE Int. Conf. Acoust., Speech, Signal Process.}, 2023, pp. 1--5.

\bibitem{chollet2017xception}
F.~Chollet, ``Xception: Deep learning with depthwise separable convolutions,'' in \emph{Proc. IEEE/CVF Conf. Comput. Vis. Pattern Recognit.}, 2017, pp. 1251--1258.

\bibitem{xu2022sjtu}
X.~Xu, Z.~Xie, M.~Wu, and K.~Yu, ``The {SJTU} system for {DCASE2022} challenge task 6: Audio captioning with audio-text retrieval pre-training,'' DCASE2022 Challenge, Tech. Rep., 2022.

\bibitem{oord2018representation}
A.~v.~d. Oord, Y.~Li, and O.~Vinyals, ``Representation learning with contrastive predictive coding,'' \emph{arXiv preprint arXiv:1807.03748}, 2018.

\bibitem{drossos2020clotho}
K.~Drossos, S.~Lipping, and T.~Virtanen, ``Clotho: An audio captioning dataset,'' in \emph{Proc. IEEE Int. Conf. Acoust., Speech, Signal Process.}, 2020, pp. 736--740.

\bibitem{kim2019audiocaps}
C.~D. Kim, B.~Kim, H.~Lee, and G.~Kim, ``Audiocaps: Generating captions for audios in the wild,'' in \emph{Proc. Conf. North Amer. Chapter Assoc. Comput. Linguistics: Hum. Lang. Technol}, 2019, pp. 119--132.

\bibitem{gemmeke2017audio}
J.~F. Gemmeke, D.~P. Ellis, D.~Freedman, A.~Jansen, W.~Lawrence, R.~C. Moore, M.~Plakal, and M.~Ritter, ``Audio set: An ontology and human-labeled dataset for audio events,'' in \emph{Proc. IEEE Int. Conf. Acoust., Speech, Signal Process.}, 2017, pp. 776--780.

\bibitem{font2013freesound}
F.~Font, G.~Roma, and X.~Serra, ``Freesound technical demo,'' in \emph{Proc. ACM Int. Conf. Multimedia}, 2013, pp. 411--412.

\bibitem{xu2023beyond}
X.~Xu, Z.~Xie, M.~Wu, and K.~Yu, ``Beyond the status quo: A contemporary survey of advances and challenges in audio captioning,'' \emph{IEEE/ACM Trans. Audio, Speech Lang. Process.}, 2023.

\bibitem{su2022one}
H.~Su, W.~Shi, J.~Kasai, Y.~Wang, Y.~Hu, M.~Ostendorf, W.-t. Yih, N.~A. Smith, L.~Zettlemoyer, and T.~Yu, ``One embedder, any task: Instruction-finetuned text embeddings,'' \emph{arXiv preprint arXiv:2212.09741}, 2022.

\end{thebibliography}

\end{document}